\documentclass[fleqn,11pt]{wlscirep}
\usepackage{epsfig}
\usepackage{longtable}
\usepackage{amsmath}
\usepackage{amsfonts}
\usepackage{listings}
\usepackage{soul,color}

\newcommand{\ket}[1]{|#1\rangle}

\newcommand{\p}{^\prime}
\newcommand{\pp}{^{\prime\prime}}

\title{Simulating electric field interactions with polar molecules using spectroscopic databases}

\author[1,*]{Alec Owens}
\author[1]{Emil J. Zak}
\author[1]{Katy L. Chubb}
\author[1]{Sergei N. Yurchenko}
\author[1]{Jonathan Tennyson}
\author[2,*]{Andrey Yachmenev}
\affil[1]{Department of Physics and Astronomy, University College London, Gower Street, London, WC1E 6BT, UK}
\affil[2]{Center for Free-Electron Laser Science (CFEL), DESY, Notkestrasse 85, 22607 Hamburg, Germany}

\affil[*]{alec.owens.13@ucl.ac.uk}
\affil[*]{andrey.yachmenev@cfel.de}

\begin{abstract}
Ro-vibrational Stark-associated phenomena of small polyatomic molecules are modelled using extensive spectroscopic data generated as part of the ExoMol project. The external field Hamiltonian is built from the computed ro-vibrational line list of the molecule in question. The Hamiltonian we propose is general and suitable for any polar molecule in the presence of an electric field. By exploiting precomputed data, the often prohibitively expensive computations associated with high accuracy simulations of molecule-field interactions are avoided. Applications to strong terahertz field-induced ro-vibrational dynamics of PH$_3$ and NH$_3$, and spontaneous emission data for optoelectrical Sisyphus cooling of H$_2$CO and CH$_3$Cl are discussed.
\end{abstract}

\begin{document}

\maketitle

\section*{Introduction}\label{sec:intro}

Molecule-field interactions are central to much current research in
molecular physics. New methods are actively being developed to manipulate the internal and external degrees of freedom, ro-vibrational state populations and coherences of molecules~\cite{Lemeshko13}. Although several experimental techniques have emerged for controlling molecular dynamics, the variety of molecules to which they have been applied to is limited. Interest in a broader range of larger systems~\cite{C4FD90017K}
along with the increasing precision of instruments has created a
demand for accurate theoretical approaches.
At present, with a few exceptions~\cite{Luis09,Reis14,Coudert15,Yachmenev16}, molecule-field interactions are described by simple minimal-coupling theoretical models (e.g. rigid rotor, harmonic oscillator). More sophisticated treatments which consider all major electronic, nuclear motion, and external field effects are therefore highly desirable.

Progress in this direction is being made but very few general purpose
computer programs exist for quantum mechanical modelling of
molecule-field interactions. The multiconfiguration time-dependent
Hartree (MCTDH) method~\cite{Beck2000,Meyer2003,Meyer2012} is one
possible option, and a robust, variational approach,
RichMol, will soon be published. However, for high
accuracy applications, calculations can be prohibitively expensive
even for small to medium sized molecules. In this work we circumvent this issue by utilizing highly accurate data, which is
freely available from the ExoMol
database~\cite{ExoMol2012,ExoMol2016}, to model phenomena associated
with polar molecules in the presence of an electric field.

A considerable amount of work goes into creating and maintaining
spectroscopic
databases~\cite{ExoMol2012,ExoMol2016,HITRAN,Hill2016,JPL,CDMS,GEISA,Rey2016} and they contain a wealth of molecular
information. From a theoretical standpoint it can take years to
construct a comprehensive line list suitable for high-resolution
spectroscopy. The process itself employs a combination of first
principles and empirically tuned quantum mechanical
methods~\cite{WIREs2012}. Because of the enormous number of
transitions which must be considered, often billions for small,
polyatomic molecules, the task is computationally intensive and
requires extensive use of high-performance computing facilities.

The ExoMol database~\cite{ExoMol2012,ExoMol2016} contains detailed
line lists for a number of important diatomic and polyatomic molecules
which have been treated using nuclear motion programs specially
optimized for such calculations~\cite{ExoSoft2016}. Originally set up
to aid the characterisation and modelling of cool stars and
exoplanets\cite{ExoMol2012}, the data has found other uses beyond
atmospheric applications. These include the calculation of molecular
partition functions~\cite{14SoHYu.PH3}, radiative lifetimes of
ro-vibrational states~\cite{lifetimes:2016}, the study of magnetic
field effects (Zeeman splittings)~\cite{jt655}, and investigating the
sensitivity of spectral lines to a possible variation of the
proton-to-electron mass ratio~\cite{Owens:2015,Owens:2016}.

Here we present two field-dependent applications which exploit the
same high accuracy data. The two phenomena we consider are
terahertz-induced two-level coherence and field-free orientation with
application to PH$_3$ and NH$_3$, and spontaneous emission data for
Sisyphus cooling of H$_2$CO and CH$_3$Cl. The Hamiltonians required to
model these situations can be built from computed ro-vibrational
transitions which are, or will soon be in the case of CH$_3$Cl,
available from the ExoMol website (www.exomol.com). Although we
only discuss these two examples the external field Hamiltonian we
describe is suitable for a range of molecule-field interaction
applications such as double resonance spectroscopy~\cite{Schmitz15},
frequency chirp excitation techniques~\cite{Rickes00}, alignment and
orientation using static and laser
fields~\cite{Friedrich99,Filsinger09}, molecular trapping and
cooling~\cite{Meerakker12}.

\section*{Results}\label{sec:results}

\subsection*{Strong terahertz field induced ro-vibrational dynamics of PH$_3$ and NH$_3$}
\label{sec:thz}

 Recent advances in high energy laser-based terahertz (THz) technology have opened up exciting opportunities in THz coherent control, field-free orientation and alignment. Generally speaking it is considerably more challenging to create oriented, as opposed to aligned, molecular samples. Transient field-free alignment usually requires non-resonant, moderately intense laser fields to be applied several times throughout an experiment to maintain the coherent evolution of a rotational wavepacket under field-free conditions (see Ref.~\citenum{Stapelfeldt03} for a review). Alignment is achieved by exciting Raman transitions between ro-vibrational states with the same parity. For field-free orientation, both even- and odd-parity angular momentum states must be coherently excited. This has been demonstrated experimentally using non-resonant, intense two color laser pulses~\cite{De09,Frumker12}, linearly polarized pulses with $45^\circ$-skewed mutual polarization~\cite{Kitano09,Fleischer09}, an optical centrifuge~\cite{Korobenko16}, and intense THz pulses~\cite{Fleischer11,Kitano13,Egodapitiya14}.

 In contrast to non-resonant laser fields, single THz pulses contain frequency components resonant to the energy spacings between adjacent, opposite-parity rotational states with $\Delta J=\pm 1$. This gives rise to transient molecular orientation which manifests, for example, in emission of free induction decay signals~\cite{Fleischer11}. A number of theoretical and experimental studies have reported strong THz field induced molecular alignment and orientation, rotational coherence and birefringence for diatomic (e.g. CO, HF, HBr)~\cite{Lapert12,Kitano13,Kitano14} and linear triatomic (e.g. OCS)~\cite{Fleischer11,Fleischer12,Lapert12,Egodapitiya14} molecules. In this section we present theoretical simulations of THz-induced ro-vibrational dynamics for PH$_3$ and NH$_3$. Field-free alignment and orientation of phosphine is demonstrated using four time-delayed THz pulses, and by employing a combination of THz and intense non-resonant Raman pulses. For ammonia we focus on creating and manipulating the rotational coherences and populations by time-delayed THz pulse pairs.

 The time-dependent quantum simulations of THz-molecule interactions
were carried out using the electric field Hamiltonian in the dipole
approximation as given by Eq.~\eqref{eq:hamiltonian}. The
ro-vibrational energies $\epsilon_{l,J}$ and dipole matrix elements
$\bar{\mu}_A^{(l,J,m,l’,J’,m’)}$ where the polarization axis $A=X,Y,Z$, were extracted from the room temperature line lists for PH$_3$~\cite{SousaSilva13} and NH$_3$~\cite{Yurchenko11} (available
from the ExoMol database; the much larger hot line lists~\cite{jt592,Yurchenko11} are not necessary for the present applications). The ro-vibrational basis was truncated at the maximal rotational quantum number $J=20$ and the maximal vibrational quanta $v_i$ corresponding to the ground and first excited vibrational states, i.e. $v_i=0,1$. For NH$_3$, due to strong anharmonicity of the large amplitude umbrella-type vibrational
motion we extended the vibrational basis set to include $v_{\rm umb}=0,\ldots,7$. 

 For both molecules only one quantum wavepacket simulation corresponding to the initial ground ro-vibrational state was performed. For PH$_3$ this was the $v=0$, $J=0$, $k=0$ state, where the quantum number $k$ is the projection of the total angular momentum onto the molecule-fixed $z$ axis. For NH$_3$, due to nuclear spin statistics, the initial state was the $v=0^+$, $J=0$, $k=0$ state. In order to investigate the influence of temperature $T$ on alignment and orientation, simulations were carried out at $T=0$~K and $T=10$~K for PH$_3$. At $T=10$~K a total of 81 initial states are populated. Final results were therefore averaged over all quantum trajectories from the different initial states.
 
 The single-pulse THz electric field was represented by the analytical function~\cite{Lapert12}
\begin{eqnarray}\label{eq:thz_shape}
E(t) = E_0\cos^2(\pi t/\bar{\tau})\sin(2\pi\bar{\nu}t),~~~t=[-\bar{\tau}/2,\bar{\tau}/2],~~~E(t)=0~\text{otherwise},
\end{eqnarray}
where $E_0$ is the amplitude of the pulse, $\bar{\tau}$ is the pulse duration, and $\bar{\nu}$ is the central frequency. In all calculations we fixed $\bar{\tau}$ to 4~ps. For simulations of the mixed Raman-THz field-free orientation of PH$_3$, the lack of a polarizability model in the ExoMol database meant the simulations of the initial Raman excitation were carried out using a more general variational approach implemented in the TROVE~\cite{Yachmenev16} and RichMol program packages. The resulting wavepacket was then used as the initial wavefunction in subsequent THz-driven dynamics simulations, modelled using the ExoMol line list data.  Note that all fields are polarized along the $Z$ axis in calculations.

\subsubsection*{PH$_3$}

\begin{figure}
\centering
\includegraphics[scale=0.55]{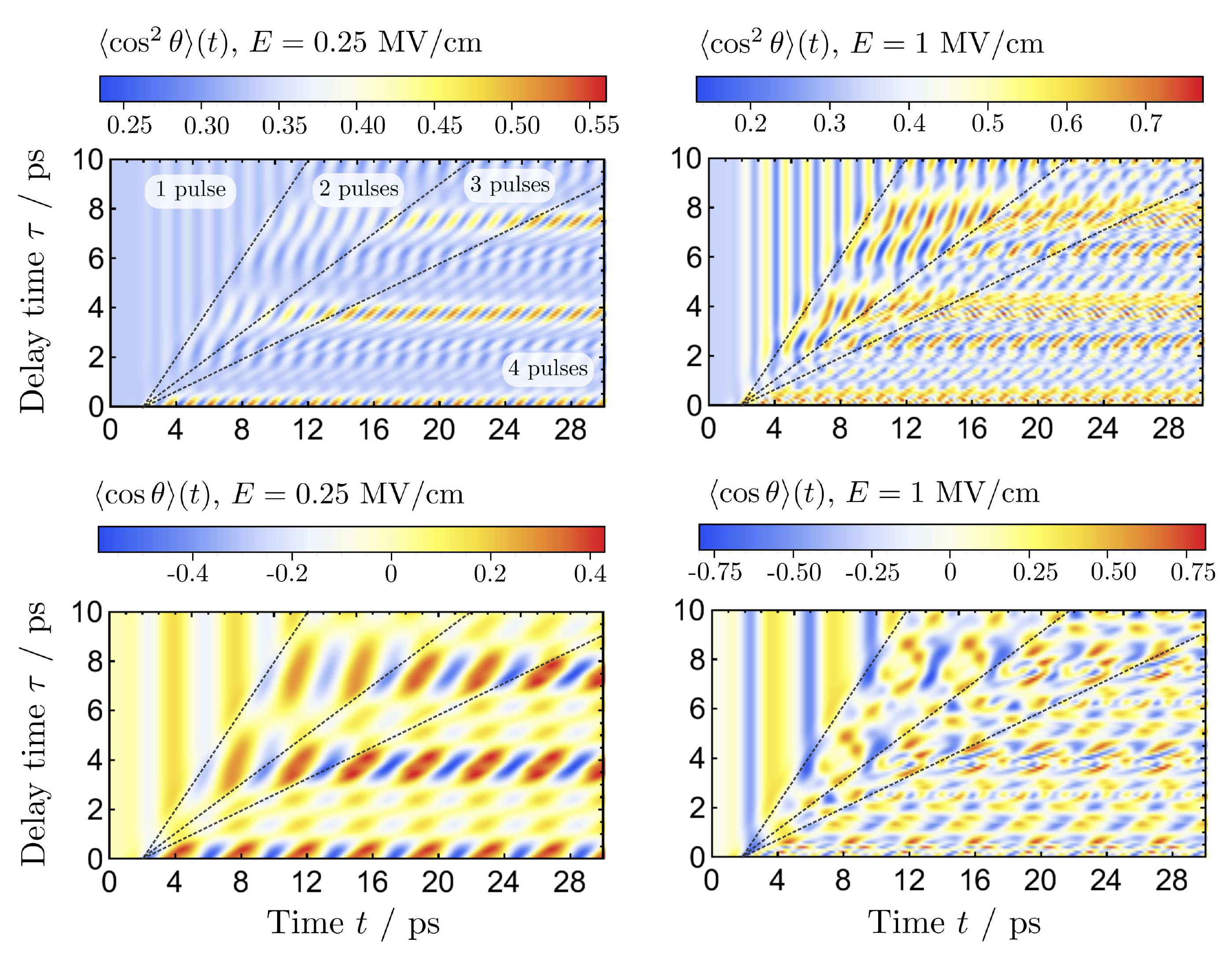}%
\caption{\label{fig:thz_ph3_0K}Alignment and orientation of PH$_3$ initially in the ground state  at $T=0$~K induced by a sequence of four time-delayed THz pulses (with central frequency $\bar{\nu}=0.5$~ps$^{-1}$) plotted for different time delays $\tau$ between the pulses. Left and right panels correspond to electric field strength $E_0$ of the THz pulse equal to 250~kV/cm and 1~MV/cm, respectively.
}
\end{figure}

\begin{figure}
\centering
\includegraphics[scale=0.55]{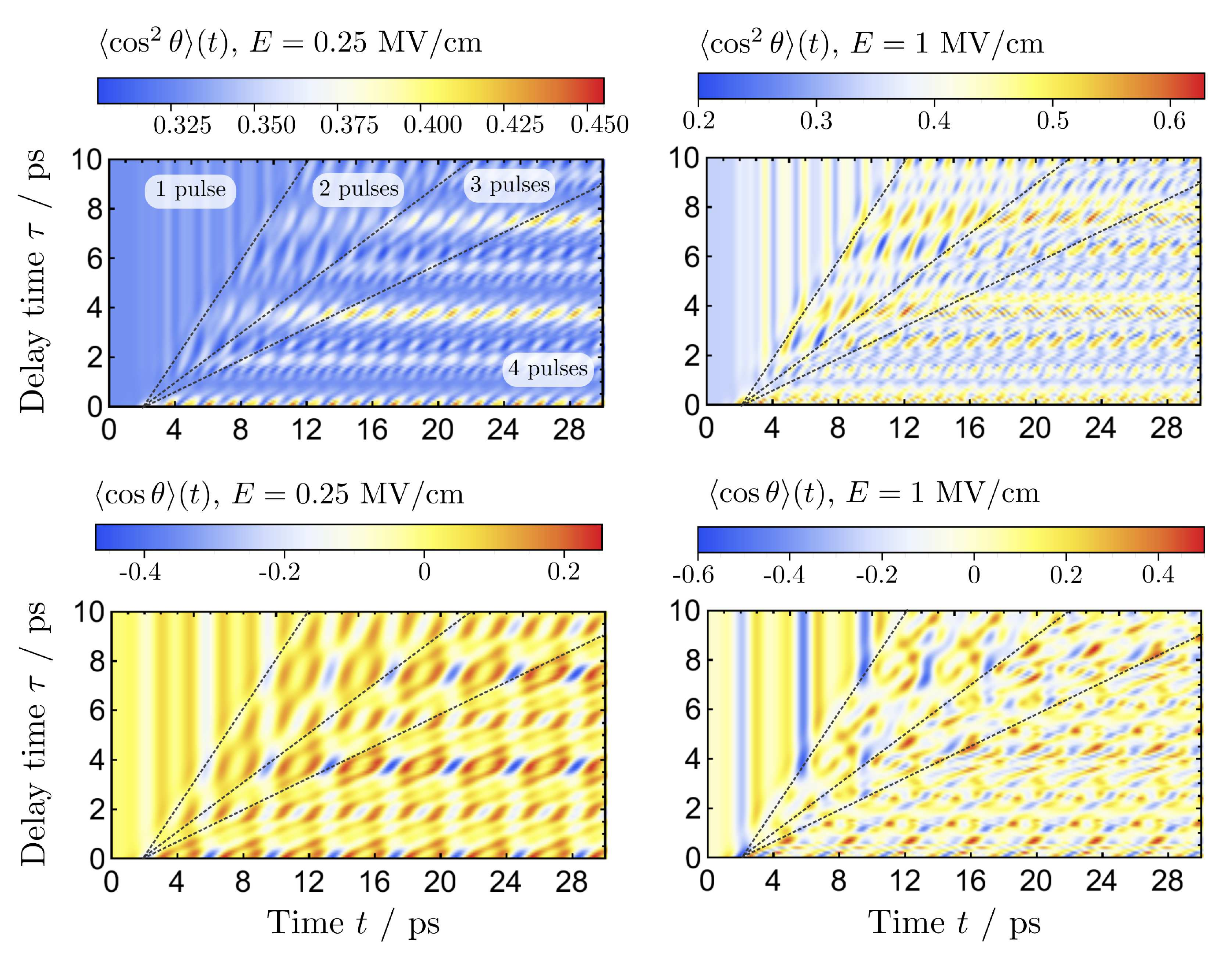}%
\caption{\label{fig:thz_ph3_10K}Alignment and orientation of PH$_3$ initially in thermal equilibrium at $T=10$~K induced by a sequence of four time-delayed THz pulses (with central frequency $\bar{\nu}=0.5$~ps$^{-1}$) plotted for different time delays $\tau$ between the pulses. Left and right panels correspond to electric field strength $E_0$ of the THz pulse equal to 250~kV/cm and 1~MV/cm, respectively.
}
\end{figure}

The results of our numerical simulations of PH$_3$ at $T=0$~K and $T=10$~K are shown in Fig.~\ref{fig:thz_ph3_0K} and Fig.~\ref{fig:thz_ph3_10K}, respectively. Four time-delayed THz pulses with central frequency $\bar{\nu}=0.5$~ps$^{-1}$ and peak field strengths at $E_0=250$~kV/cm and $E_0=1$~MV/cm have been used to induce transient alignment $\langle\cos^2\theta\rangle(t)$, and orientation $\langle\cos\theta\rangle(t)$, where $\theta$ is the Euler angle. Such intense THz pulses in the few-cycle regime can be routinely generated in a convenient tabletop setup~\cite{Yeh07,Stepanov08,Hirori11}. As expected, increasing the number of pulses significantly enhances the degree of orientation and alignment for both temperatures and peak field strengths. The periodic behaviour of the orientation dynamics follows the quantum rotational revival pattern; strong peaks are observed at times $t\approx 2.0, 4.2, 5.7, 7.9, 9.5$~ps etc., which is approximately $1/2, 1, 3/2, 2, 5/2,\ldots$ of a revival time $T_{\rm rev}= 1/2Bc \approx 3.7$~ps ($B\approx 4.5$~cm$^{-1}$ is the rotational constant of PH$_3$ and $c$ is the speed of light). The enhancement of the alignment and orientation at delay times $\tau\approx 4, 8, 12$~ps etc., reveals recurrences separated by approximately $\Delta\tau =T_{\rm rev}$.

 Increasing the peak field strength of the THz pulses results in a more complex pattern of alignment and orientation with additional local minima and maxima. These secondary peaks are always present but attenuated in the weak field regime. Alignment and orientation are enhanced in stronger electric fields, hence higher absolute peak values of $\langle\cos^2\theta\rangle(t)$ and $\langle\cos\theta\rangle(t)$. However, raising the temperature from $T=0$~K to $T=10$~K activates several new quantum trajectories via thermally populated states and this leads to overall decoherence of the wavepacket. This effect is responsible for the lower absolute values of the orientation and alignment parameters at $T=10$~K in Fig.~\ref{fig:thz_ph3_10K}. Therefore, losses in orientation or alignment at higher temperatures can be compensated for with a stronger peak intensity of the THz pulse.

\begin{figure}
\centering
\includegraphics[scale=0.25]{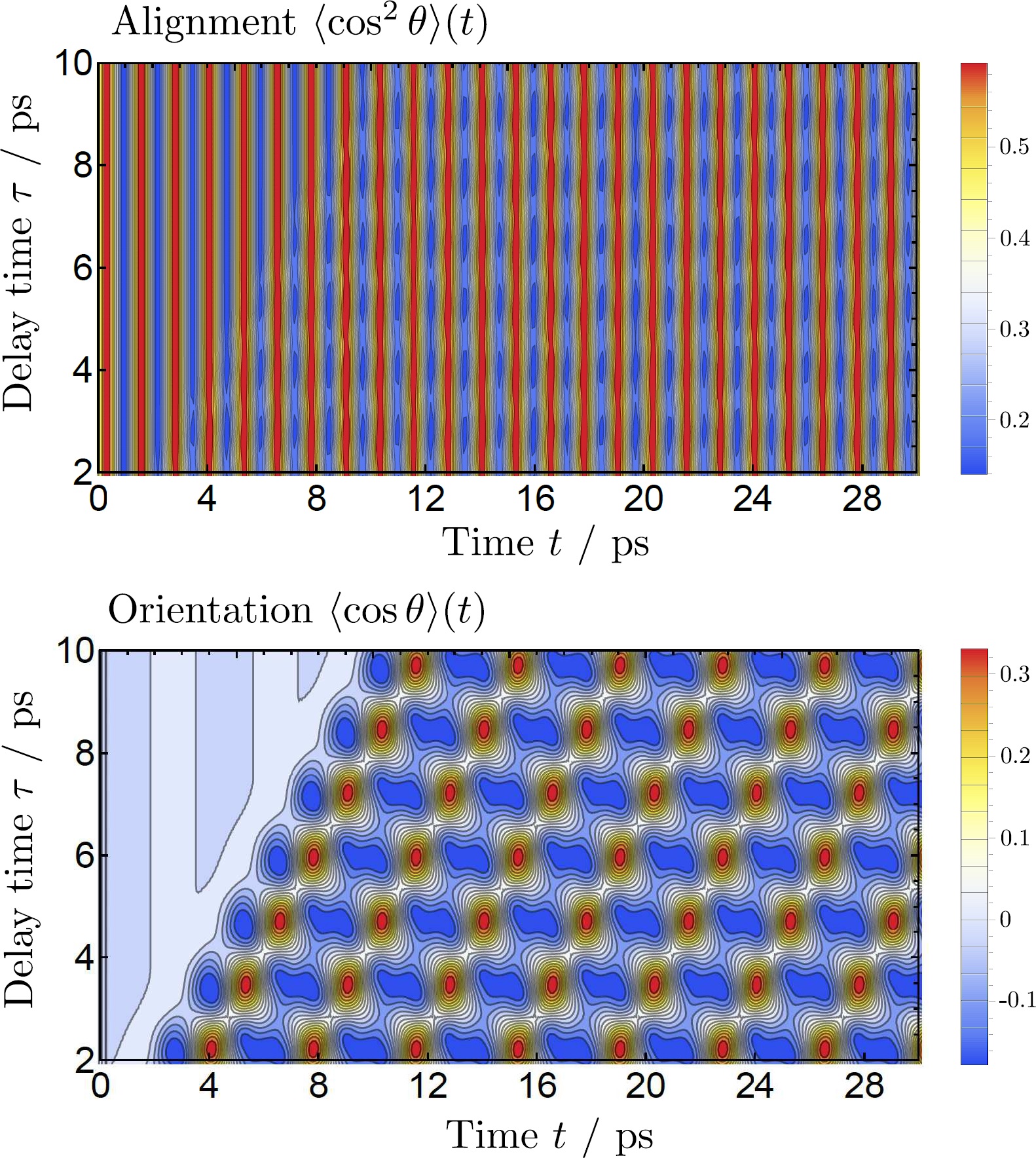}%
\caption{\label{fig:raman_thz_ph3}Alignment and orientation of PH$_3$ initially in the ground state induced by a time-delayed Raman Gaussian pulse ($E_0=10^8$~V/cm, $\omega=800$~nm, ${\rm FWHM}=0.4$~ps) and THz pulse ($\bar{\nu}=0.5$~ps$^{-1}$, $E_0=250$~kV/cm) plotted for different time delays $\tau$ between the pulses.
}
\end{figure}

 As an alternative to multiple THz pulses, in Ref.~\citenum{Egodapitiya14} it was demonstrated that the degree of alignment and orientation can be improved by applying a short intense Raman pulse to coherently prepare molecules in highly excited rotational states prior to THz exposure. This allows a larger molecular population to occupy states for which transition frequencies match the peak of the THz pulse envelope. Raman excitation follows $\Delta J=\pm 2$ selection rules and hence couples states of the same parity. Therefore, when starting from an initial $J=0$ state, the Raman pulse predominantly excites $J=2$ states. The subsequent THz pulse induces $J\rightarrow J': 0\rightarrow 1, 2\rightarrow 3$ transitions and produces a mixed-parity wavepacket, thus enabling effective net orientation. The results of our simulations for PH$_3$ are shown in Fig.~\ref{fig:raman_thz_ph3}. The alignment pattern is insensitive to the time delay between the Raman and THz pulse and it appears the THz pulse contributes only marginally to alignment. For a given time delay the orientation peaks are separated by $\approx 3.9$~ps, which corresponds to $T_{\rm rev}$.

\subsubsection*{NH$_3$}

 As a second example we explore strategies to establish control over rotational coherences and populations in NH$_3$ by applying optimally timed THz pulses. Recently it was shown both theoretically and experimentally~\cite{Fleischer12} that a pair of THz pulses separated in time can induce larger population transfer (larger transient emission responses) than two interactions within a single short THz pulse.

\begin{figure}
\centering
\includegraphics[width=\textwidth]{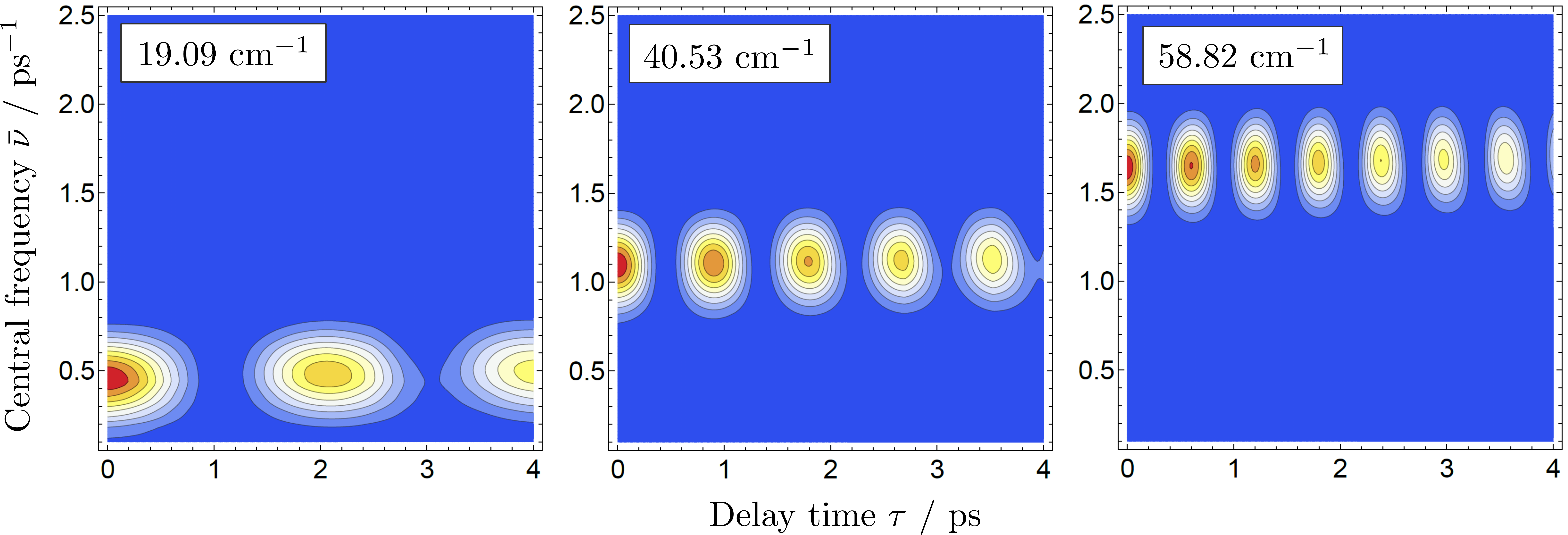}%
\caption{\label{fig:two_thz_fourier}Fourier intensity of two THz pulses separated by a delay time $\tau$ versus central frequency $\bar{\nu}$ of a single pulse (see Eq.~\eqref{eq:thz_shape}). Results are computed at characteristic frequencies 19.09, 40.53, and 58.82 cm$^{-1}$ which correspond to rotational transition frequencies of NH$_3$ for the $|0,1\rangle\rightarrow|1,0\rangle$, $|1,0\rangle\rightarrow|2,1\rangle$, and $|2,1\rangle\rightarrow|3,0\rangle$ transitions, respectively. States are labelled as $|J,v_{\rm umb}\rangle$.
}
\end{figure}

  To adjust the THz pulse parameters to the energy level structure of NH$_3$, we first look at the Fourier intensity of a pair of THz pulses at a few characteristic frequencies as a function of the time delay $\tau$ between the pulses and the pulse central frequency $\bar{\nu}$ (see Eq.~\eqref{eq:thz_shape}). The results are shown in Fig.~\ref{fig:two_thz_fourier} for the three rotational transitions, $|0,1\rangle\rightarrow|1,0\rangle$, $|1,0\rangle\rightarrow|2,1\rangle$, and $|2,1\rangle\rightarrow|3,0\rangle$, where states of NH$_3$ are labelled as $|J,v_{\rm umb}\rangle$, and the quantum numbers $k$ and $m$, which correspond to the projection of the total angular momentum onto the molecule-fixed and laboratory-fixed axes, respectively, are zero. It is evident that the optimal values of $\bar{\nu}$ are 0.4, 1.1, and 1.6~ps$^{-1}$ for the three different rotational excitations of NH$_3$.

\begin{figure}
\centering
\includegraphics[scale=0.25]{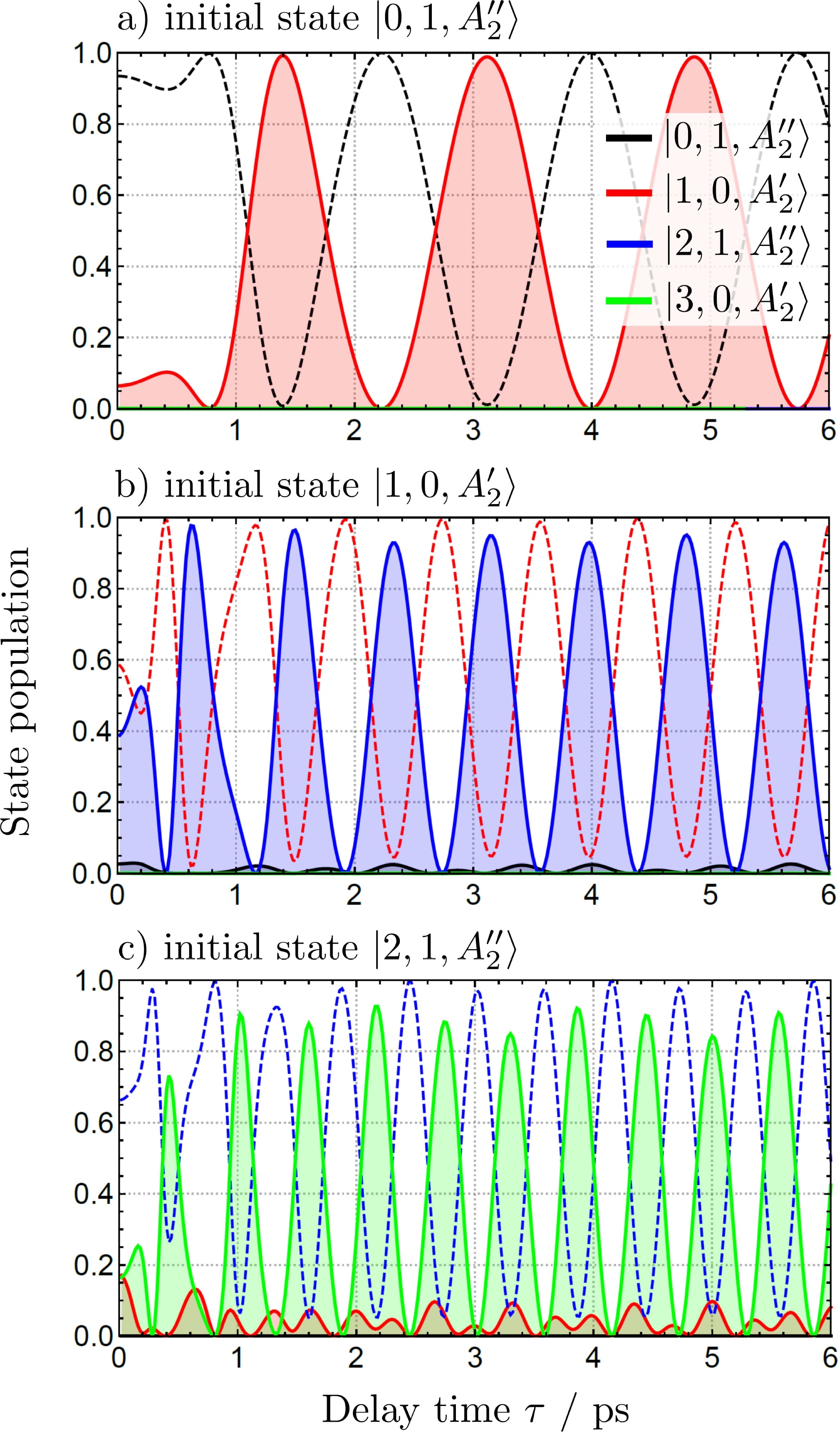}%
\caption{\label{fig:two_thz_nh3}Ro-vibrational wave packet populations of NH$_3$ after excitation using a pair of THz pulses with $E_0=500$~kV/cm, $\bar{\tau}=4$~ps, (a) $\bar{\nu}=0.4$~ps$^{-1}$, (b) $\bar{\nu}=1.1$~ps$^{-1}$, (c) $\bar{\nu}=1.6$~ps$^{-1}$, as a function of delay time $\tau$ between the pulses. The different panels, (a)--(c), correspond to an initial wave packet in the ground or excited ro-vibrational states of NH$_3$ ($A_2''$ and $A_2'$ denote state symmetry in {\bf D}$_{\rm 3h}$(M)). Population of the initial state is plotted with a dashed line.
}
\end{figure}

 Fig.~\ref{fig:two_thz_nh3} depicts the populations of the ro-vibrational wave packets of NH$_3$, initially in the $|0,1\rangle$, $|1,0\rangle$, and $|2,1\rangle$ ro-vibrational states, following the excitation by a time-delayed THz pulse pair with $E_0=500$~kV/cm and respective pulse central frequencies $\bar{\nu}=0.4$, 1.1, and 1.6~ps$^{-1}$. Population transfer is significantly enhanced by two properly delayed THz pulses when compared to a single, but twice as intense, THz pulse ($\tau=0)$. Slight adjustment of the central frequency of the THz pulses increases the selectivity of excitation, suppressing (or enhancing) two-photon transitions. As expected, population modulations for the three excited rotational states with $J=1$, 2 and 3, shown in the (a)--(c) panels of Fig.~\ref{fig:two_thz_nh3}, appear with a respective revival period of $T_{\rm rev}\approx J/2Bc \approx1.7$, 0.8 and 0.6~ps, respectively, where $B\approx6.35$~cm$^{-1}$ is the rotational constant of NH$_3$.

\begin{figure}
\centering
\includegraphics[scale=0.25]{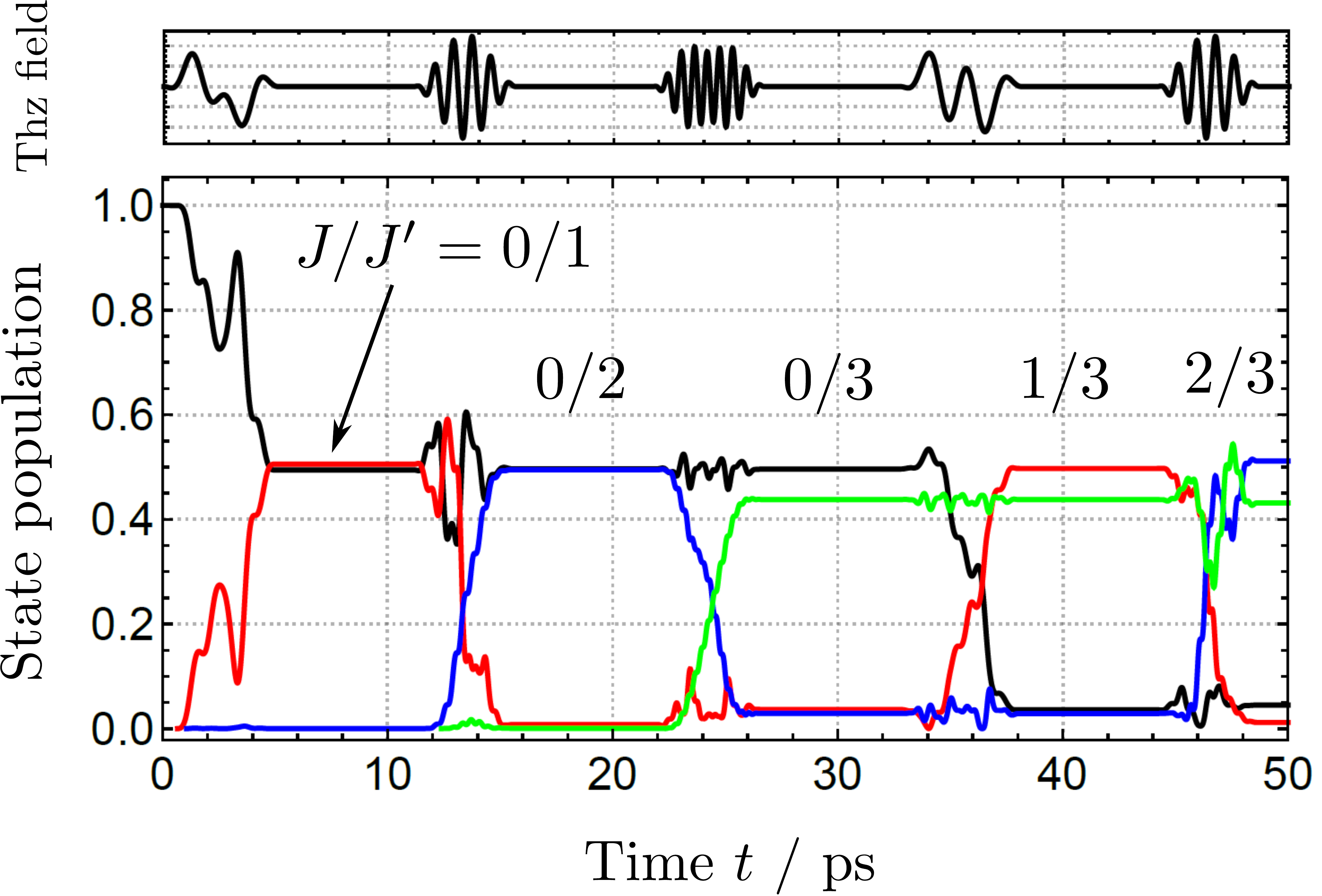}%
\caption{\label{fig:train_thz_nh3}Population transfer between pairs ($J/J'$) of rotational states of NH$_3$ driven by a sequence of five time-delayed THz pulse pairs. Each pulse pair is separated in time by 10~ps. The field parameters $(\bar{\nu}~{\rm ps}^{-1},\tau~{\rm ps})$ for each pulse pair are: $(0.4,1.1)$, $(1.1,0.6)$, $(1.6,1.0)$, $(0.4,1.4)$, $(1.1,0.6)$. The color scheme is defined in Fig.~\ref{fig:two_thz_nh3}.
}
\end{figure}

 Based on the results of Fig.~\ref{fig:two_thz_nh3} we can combine pairs of THz pulses to create coherences between different pairs of rotational states with even and odd parities. This is shown in Fig.~\ref{fig:train_thz_nh3}. The first pulse at $t\approx 1.1$~ps induces a $J\rightarrow J': 0\rightarrow 1$ transition, creating a coherent superposition  $\frac{1}{\sqrt{2}}\left(|0,1\rangle \pm |1,0\rangle \right)$. The second pulse is tuned to affect population of the $|1,0\rangle$ state and transforms the two-state coherence into a new one:  $\frac{1}{\sqrt{2}}\left(|0,1\rangle \pm |2,1\rangle \right)$. Parameters of the subsequent pulses (upper inset in Fig.~\ref{fig:train_thz_nh3}) are optimized to maintain uniform division of molecular population between the two states. This illustrates the high level of control which can be exerted over the ro-vibrational wavepacket with the use of THz pulses.

\begin{figure}
\centering
\includegraphics[width=\textwidth]{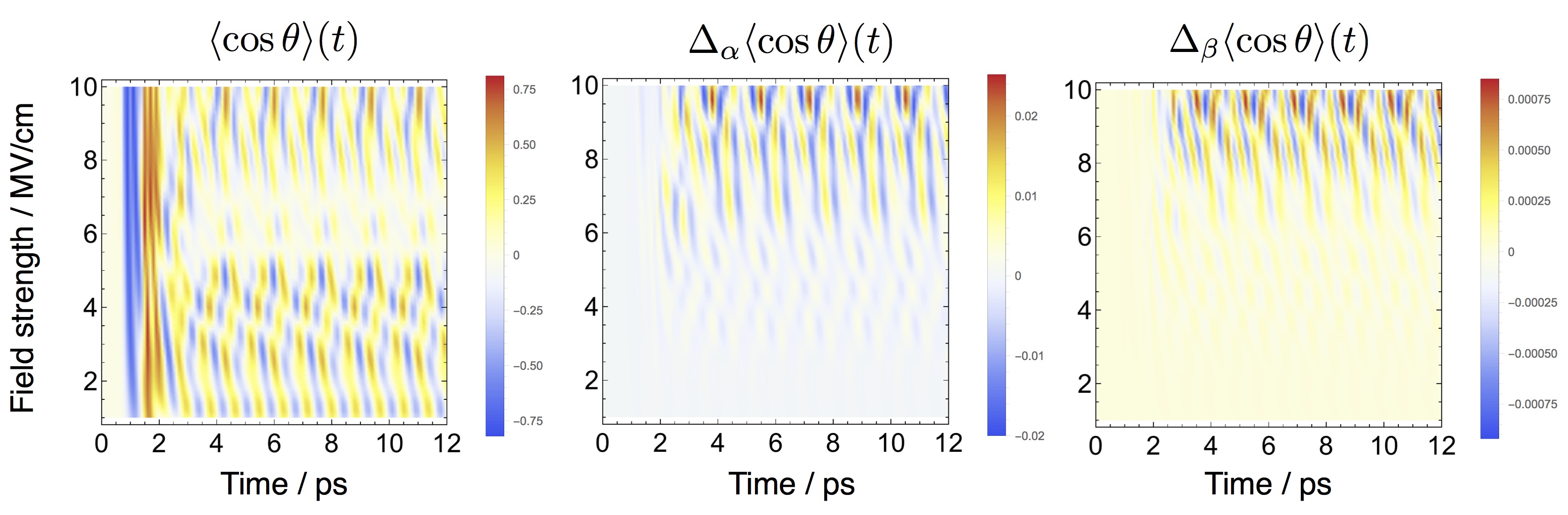}%
\caption{\label{fig:orientation_thz_nh3}Orientation of NH$_3$ driven by THz pulses with electric field strengths $E_0$ up to $10$~MV/cm range. The left panel corresponds to the electric field Hamiltonian in the dipole approximation. The middle panel shows the difference in orientation when the polarizability $\alpha$ is included in the Hamiltonian. The right panel shows the difference in orientation when the hyperpolarizability $\beta$ is also included.}
\end{figure}

 To further investigate the validity of our approach we have computed the alignment and orientation of NH$_3$ using three different Hamiltonian models. The first, given in Eq.~\eqref{eq:hamiltonian}, is the electric field Hamiltonian in the dipole approximation. The second Hamiltonian we use incorporates the polarizability $\alpha$ into the model, while the third Hamiltonian also includes the hyperpolarizability $\beta$. Latter calculations were perfomed using our recently developed variational approach RichMol. The THz field was composed of three simultaneously applied pulses with respective central frequencies $\bar{\nu}=0.3$, 0.6, and 0.9~ps$^{-1}$ each with duration $\bar{\tau}=4$~ps. The results are illustrated in Fig. \ref{fig:orientation_thz_nh3} for electric fields strengths up to $10$~MV/cm. Calculations were performed on an initial ground vibrational state with $J=1$, $k=1$, $m=0$ and mixed parity $+/-$; that is $|\psi\rangle = \frac{1}{\sqrt{2}}(|1,1,0,0^+\rangle+|1,1,0,0^-\rangle)$.
 
 As shown in Fig.~\ref{fig:orientation_thz_nh3} it is only for the very intense $10$~MV/cm THz pulses that we see contributions from the polarizability and hyperpolarizability, otherwise their effects are negligible. In general, alignment and orientation caused by the hyperpolarizability is an order of magnitude smaller than the polarizability and contributes with opposite sign. Field strengths as strong as $10$~MV/cm, where the non-linear THz effects of $\alpha$ and $\beta$, although noticeable, are small enough to be neglected, can therefore be modelled using the Hamiltonian in the dipole approximation in conjunction with ExoMol line list data.

\begin{figure}
\centering
\includegraphics[width=\textwidth]{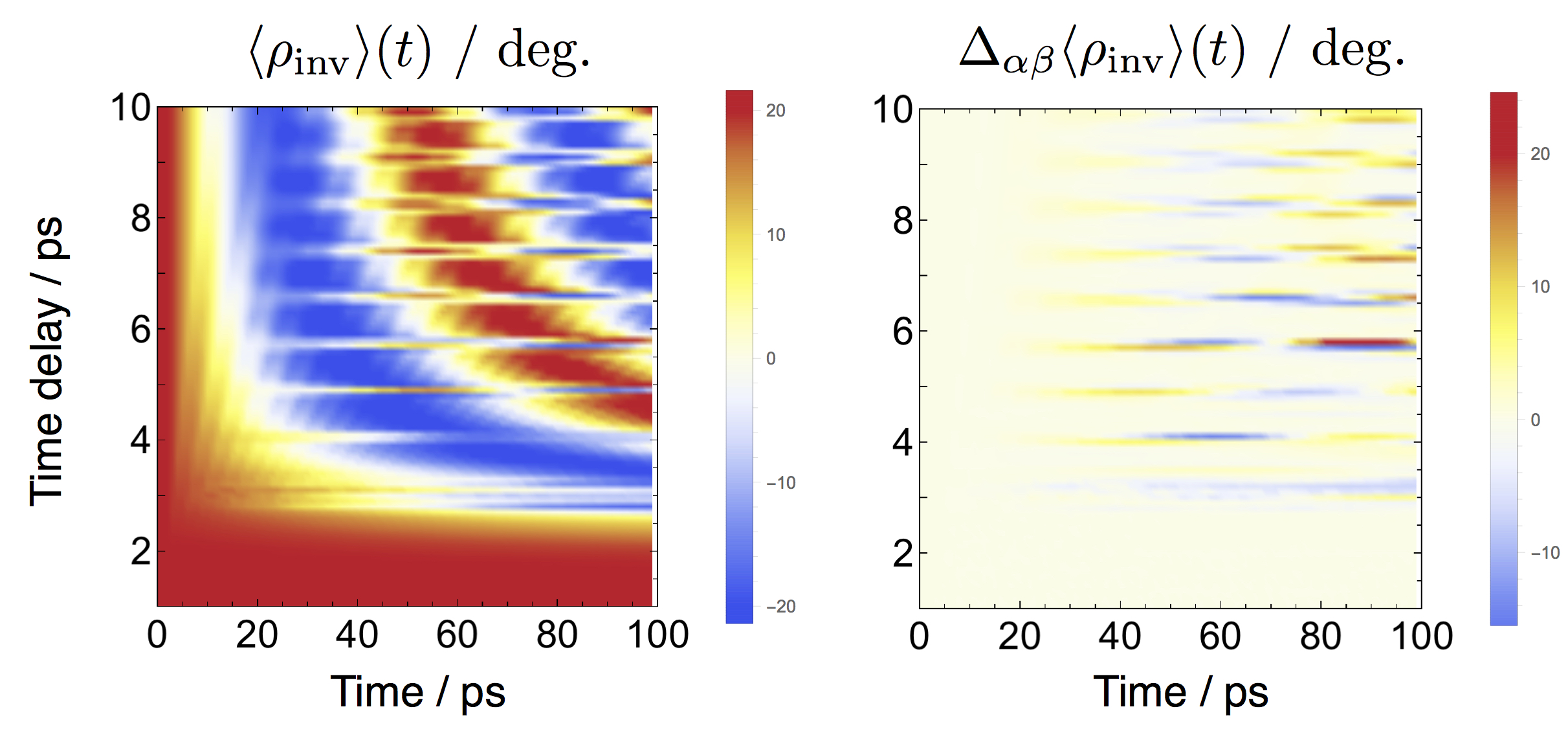}%
\caption{\label{fig:tunnelling_thz_nh3}The effect of intense quasi half-cycle THz pulses ($E_0=50$~MV/cm, $\bar{\nu}=0.1$~ps$^{-1}$, $\bar{\tau}=4$~ps) on the dynamics of the large-amplitude vibrational coordinate $\rho_{\rm inv}$ of NH$_3$. The left panel shows the expectation value of $\rho_{\rm inv}$ (zero at planar molecular geometry) as a function of time and delay time $\tau$ between single pulses. The right panel shows the 
contribution of the polarizability $\alpha$ and hyperpolarizability $\beta$ when included in the model.}
\end{figure}

Intense THz pulses offer a promising approach for controlling large-amplitude motions in floppy molecules. In Fig.~\ref{fig:tunnelling_thz_nh3} we investigate the effect of an intense THz-pulse-train on the dynamics of the large-amplitude vibrational coordinate of NH$_3$ associated with the umbrella inversional motion. This was performed with 50~MV/cm intense quasi half-cycle THz pulses, obtained in Eq.~\eqref{eq:thz_shape} by setting the single-pulse central frequency $\bar{\nu}=0.1$~ps$^{-1}$ and pulse duration $\bar{\tau}=4$~ps. The left panel of Fig.~\ref{fig:tunnelling_thz_nh3} shows the expectation value of the inversion coordinate $\rho_{\rm inv}$ (zero at planar molecular geometry) as a function of time and delay time $\tau$ between single pulses. In the region where the delay time is smaller than the single pulse duration, i.e. $\tau\leq\bar{\tau}=4$~ps, the inversion tunnelling is significantly extended or prohibited by the effects of the strong electric field. For many experiments on controlled molecules the presence of an intense electric field can lead to undesirable side effects. To avoid them, the delay time between THz pulses must exceed the single pulse duration, $\tau>\bar{\tau}$, creating short windows in time with no strong electric field present.
As seen in Fig.~\ref{fig:tunnelling_thz_nh3}, for delay times just above the pulse duration it is possible to modulate, at least to some extent, the inversion tunnelling rate. For larger delay times the tunnelling dynamics quickly approach field-free behaviour.

 To again validate the use of ExoMol line list data with the Hamiltonian in the dipole approximation for this type of observable, we have run calculations with an extended Hamiltonian containing non-linear contributions from the polarizability and hyperpolarizability tensors. The results are shown in the right panel of Fig.~\ref{fig:tunnelling_thz_nh3} and confirm the validity of the ExoMol dipole-only approach to predict non-rigid dynamics with THz fields below 50~MV/cm.

\subsection*{Spontaneous emission data for Sisyphus cooling of H$_2$CO and CH$_3$Cl}
\label{sec:sis}
Optoelectrical Sisyphus cooling is a robust and general method for producing ultracold polyatomic
molecules~\cite{Zeppenfeld:2009,Zeppenfeld:2012,Prehn:2016}. This
technique works by moving molecules through a closed system of
trapped ro-vibrational states in the presence of an electric field.
Molecules lose kinetic energy as they travel up the electric field
gradient of the trapping potential. The cycle is repeated until
sufficient cooling has taken place with sub-millikelvin temperatures
possible. The speed and efficiency of Sisyphus cooling is dictated by the Einstein A coefficients and decay channels of the involved ro-vibrational energy levels; information which is readily available and computed to a high degree of accuracy in the ExoMol database. It is possible then to identify suitable transitions in different molecular systems which could be used for Sisyphus cooling. This type of analysis can aid future experiments if there is interest in a particular molecule, potentially leading to improved rates of cooling.

\begin{figure}
\centering
\includegraphics{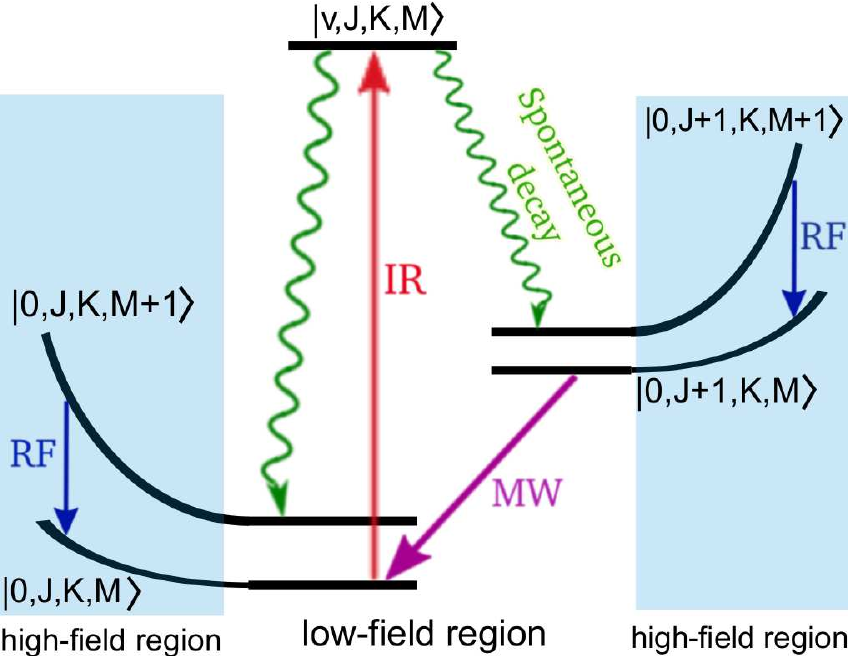}
\caption{\label{fig:sis}Overview of Sisyphus cooling scheme.}
\end{figure}

 A general overview of the Sisyphus cooling level scheme is shown in Fig.~\ref{fig:sis}. Starting from an initial, preferably highly populated ro-vibrational state, molecular population is transferred to an excited bridge state via an infrared (IR) transition. Spontaneous decay from the bridge state should be dominated by at most two channels to long-lived low-field seeking states. Large branching ratios are essential in this step with the Einstein A coefficient of the primary decay channel dictating the speed of the cooling cycle. As molecules penetrate the high-field region they climb up the Stark induced energy gradient and lose kinetic energy. For larger values of the quantum number $M$, the gradient is steeper and more energy is removed. It is important that the Stark split $M$ sublevels are low-field seeking states (shifted towards higher energies) as this ensures the molecules remain trapped. Applied radio-frequency (RF) radiation stimulates emission to lower $M$ quantum number states and the molecules gain a small amount of kinetic energy as they move down the electric field gradient. However, this gain is smaller than the previous loss in energy and the temperature of the molecules is reduced. A microwave (MW) field connects the two decay channels and closes the cycle. The process is repeated until sufficient cooling has occurred.

 Sisyphus cooling has been realised experimentally for methyl fluoride
 (CH$_3$F)~\cite{Zeppenfeld:2012} and formaldehyde
 (H$_2$CO)~\cite{Prehn:2016}. However, a number of symmetric top
 molecules with strong parallel vibrational transitions have been
 proposed as suitable candidates for cooling using this
 method~\cite{Zeppenfeld:2009}. The closed system of ro-vibrational
 energy levels should satisfy the following criteria:
\begin{itemize}
\item The initial state is highly populated and long-lived. Energy levels in the vibrational ground state with rotational quantum number $J$ in the range $2\leq J\leq 4$ are generally good candidates.
\item There must be a sufficiently fast (intense) IR transition from the initial state to an excited bridge state with an Einstein A coefficient large enough that population transfer can occur in the time frame of effective trapping.
\item State-selective spontaneous decay is possible from the bridge state with one or two dominant decay channels. The branching ratio of the two channels should sum to near unity.
\item Target states of spontaneous decay must be low-field seeking states with sufficiently long lifetimes to ensure the molecules remain trapped.
\item RF transitions from target states to lower $M$ quantum number states are available.
\item MW radiation can be used to move population from the secondary decay channel back to the initial state to close the cycle.
\end{itemize}

To illustrate how the ExoMol database, or in general any reasonably
``complete'' spectroscopic library, can be used to identify molecular
transitions for Sisyphus cooling we have performed analysis on
H$_2$CO and CH$_3$Cl. These systems have been treated within the
ExoMol framework~\cite{ExoSoft2016} and we refer the reader to the
relevant publications for details of the ro-vibrational
calculations~\cite{YaYuJe11.H2CO,15AlYaTe.H2CO,15OwYuYa.CH3Cl,16OwYuYa.CH3Cl}. Candidate ro-vibrational states were identified by sorting the \texttt{.trans} files according to upper state energy. There are a huge number of energy levels to process so to reduce the workload we only considered states involved in at least one intense transition (Einstein A coefficient of the order $10^1$ or greater) and with $J\leq 4$. Once identified, all transitions to the upper state were collected and branching ratios computed. Energy levels with two dominant decay channels or less were then analysed to check that the lower states (or target states) were low-field seeking.

\subsubsection*{H$_2$CO}

 In the work of Ref.~\citenum{Prehn:2016}, Sisyphus cooling of formaldehyde utilized ro-vibrational energy levels of the $\nu_1$ symmetric C{--}H stretching mode, $\ket{\nu_1,3,3}$, and of the vibrational ground state, $\ket{0,3,3}$ and $\ket{0,4,3}$. Here energy levels are labelled as $\ket{\mathrm{Mode},J,K}$ where $K=|K_A|$ and we omit the quantum number $K_C$ (see discussion in Ref.~\citenum{Prehn:2016}). The primary decay channel $\ket{\nu_1,3,3}\rightarrow\ket{0,3,3}$ had a branching ratio of $\gamma=0.75$ and an estimated spontaneous decay rate (equivalent to the Einstein A coefficient) of approximately $60\,$Hz, which determines the speed of cooling.

 In Table~\ref{tab:h2co} we list six collections of suitable states for cooling including those used by Ref.~\citenum{Prehn:2016} (third and forth row). Using the ExoMol database we found that the $\ket{\nu_1,3,3}$ level has 41 decay channels, of which only transitions to the $\ket{0,3,3}$ and $\ket{0,4,3}$ levels were dominant. We predict a slightly smaller Einstein A coefficient for the $\ket{\nu_1,3,3}\rightarrow\ket{0,3,3}$ transition and hence a slightly slower cooling rate than Ref.~\citenum{Prehn:2016}. However, the difference is small ($\approx10\,$Hz) and our computed branching ratios are identical to those of Ref.~\citenum{Prehn:2016}.

 For other possible candidate states in the $\nu_1$ manifold, a slightly higher branching ratio and Einstein A coefficient are predicted for the $\ket{\nu_1,4,4}\rightarrow\ket{0,4,4}$ transition. Alternatively, the $\nu_2$ C{--}O stretching mode appears suitable for Sisyphus cooling but using these states would half the speed of cooling compared to the $\nu_1$ levels. All lower states $E\pp$ in Table~\ref{tab:h2co} belong to the vibrational ground state and thus remain accessible either thermally or by methods of selective state preparation. At $T=100\,$K for example, lower states are thermally populated to around $30\%$. Note that only transitions allowed by selection rules were considered, i.e. $\Delta J$, $\Delta M=0,\pm 1$ and $\Delta K=0$.

\subsubsection*{CH$_3$Cl}

 Methyl chloride has not yet been considered for Sisyphus cooling but we expect a similar experiment to CH$_3$F~\cite{Zeppenfeld:2012} could be performed. As shown in Table~\ref{tab:ch3cl}, ro-vibrational levels of the $\nu_{1}$ symmetric CH$_{3}$ stretching mode and vibrational ground state appear the best candidates for cooling. Similar to methyl fluoride, which had a spontaneous decay rate of about $15\,$Hz for the $\ket{\nu_1,3,3}\rightarrow\ket{0,3,3}$ transition, the respective rate in CH$_3$Cl is only marginally larger and would hence provide a similar speed of cooling. Note that we have only considered CH$_3{}^{35}$Cl for the present analysis but we would expect the same conclusions for CH$_3{}^{37}$Cl.

\section*{Discussion}
\label{sec:conc}

Comprehensive spectroscopic line lists generated as part of the ExoMol project can be used to model ro-vibrational phenomena for small polyatomic molecules in the presence of external electric fields. By doing so we avoid repeating the same, often expensive calculations.
Currently the ExoMol database contains complete sets of high accuracy ro-vibrational energies, transition frequencies, Einstein A coefficients, and complex dipole moment phases for many important diatomic and small polyatomic molecules valid for temperatures up to $T=1500$~K.
These data can be straightforwardly utilized to model molecule-field  interactions in the dipole approximation.
In the future we plan to extend the ExoMol database with Raman and electric quadrupole transition moments for selected molecules, such as NH$_3$ and H$_2$O, thereby extending the range of possible applications.

Two illustrative examples of molecule-field interaction applications were presented. Strong THz field induced ro-vibrational dynamics of PH$_3$ and NH$_3$ were simulated and these represent the first high accuracy calculations on polyatomic molecules which have been reported in the literature. For optoelectrical Sisyphus cooling, suitable collections of states were identified in H$_2$CO and CH$_3$Cl. In both molecules the $\nu_1$ symmetric stretching mode provided the fastest cooling rates and most suitable ro-vibrational energy levels for Sisyphus cooling. 

 Although we have chosen relatively straightforward processes to look at, our approach can simulate any case with no constraint on the availability of data (provided the molecule is in the ExoMol database). At present, we are unaware of any other line list database which stores the dipole moment phase factors. This information is always produced with the dipole line strengths or Einstein coefficients but is discarded when the line lists are compiled, thus making them unusable for modeling field-driven effects. We therefore encourage other theory-based data compilations to retain this information in future.

\section*{Methods}
\label{sec:methods}

\subsection*{ExoMol line list data structure}
\label{sec:data}

A complete description of the ExoMol data structure along with
examples was recently reported in Ref.~\citenum{ExoMol2016}. Therefore we
provide only a brief description that is relevant for the present
work. The two main files available for download from the ExoMol
website (www.exomol.com) are the \texttt{.states} and \texttt{.trans}
files, unique to each molecule. The \texttt{.states} file contains all
computed ro-vibrational energy levels (in cm$^{-1}$). For polyatomics,
which are the focus of the present study, each energy level has a
unique state ID with symmetry and quantum number labelling. The
\texttt{.trans} files, which are split into frequency windows as to be
more manageable, contain all calculated molecular transitions. Upper
and lower state labels, transition frequencies and Einstein A
coefficients are provided.

 Modelling Stark associated phenomena requires information on the complex phase of the dipole moment matrix elements (discussed below). This information, simply a $+$ or $-$, is not currently available for all molecules in the ExoMol database. However, work is underway to rectify this and the complex phase for each molecular transition can be extracted from the \texttt{.dipole} file. Several other molecule specific files related to atmospheric applications (e.g. pressure broadening of spectral lines) are available but since we do not use them in this work, we do not discuss them here.

\subsection*{Electric field interaction Hamiltonian}
\label{sec:hamiltonian}

 The complete Hamiltonian is written as a sum of the field-free ro-vibrational Hamiltonian and the molecule-field interaction Hamiltonian, which is approximated in this work as the permanent dipole interaction
\begin{eqnarray}\label{eq:hamiltonian}
H = \sum_{l,J,m}\varepsilon_{l,J}|l,J,m\rangle \langle l,J,m| -\sum_{l,J,m,l',J',m'}\sum_{A=X,Y,Z}E_A(t)\bar{\mu}_A^{(l,J,m,l’,J’,m’)} |l',J',m'\rangle \langle l,J,m|.
\end{eqnarray}
Here $\varepsilon_{l,J}$ and  $|l,J,m\rangle$ denote the field-free ro-vibrational energy and wavefunction, respectively, of a state identified by the running number $l$ (ID number)  in the line list energy file (\texttt{.states} file), $J$ is the rotational quantum number of the total angular momentum, and the quantum number $m=-J,\ldots,J$ is the projection of the total angular momentum on the laboratory-fixed $Z$ axis. The coupling strength is given by the product of the ro-vibrational matrix elements of electronic ground state electric dipole moment $\bar{\mu}_A^{(l,J,m,l’,J’,m’)}$ in the ro-vibrational basis and the time-dependent classical electric field $E_A(t)$, defined in the laboratory fixed axes system. To describe the radiative fields we use the form $E_A(t)=E_A^{(0)}(t;t_0,T)\cos( \omega_A t)$, where $A$ is the polarization axis, $E_A^{(0)}(t;t_0,T)$ is the pulse time profile with a maximum at $t=t_0$ and duration (FWHM) $T$, and $\omega_A$ is the carrier frequency.

 Calculation of the $\bar{\mu}_A$ ro-vibrational matrix elements from Einstein A coefficients, extracted from the ExoMol line list, is described in detail below. The selection rules for $\bar{\mu}_A$ allow coupling between ro-vibrational states with $\Delta J = J'-J=0,\pm1$ and $\Delta  m = m'-m=0,\pm1$. Hence, the total Hamiltonian may be constructed in the form of a block-tridiagonal matrix, having blocks corresponding to $\Delta m =-1$, 0, or +1 in the lower, main, and upper diagonals, respectively. Using this representation, various linear algebra operations can be performed efficiently and the memory requirements reduced by use of the band matrix storage scheme.
For $A=Z$, $\bar{\mu}_Z$ can couple only states with $\Delta m=0$, thus the total Hamiltonian becomes factorized into independent blocks for each chosen $m$, which are processed independently.

 The time-evolution of the wavefunction is given by the time-dependent Schr\"odinger equation with the Hamiltonian in Eq.~\eqref{eq:hamiltonian}, which may be time-independent or time-dependent. For static fields the problem reduces to solving an eigenvalue equation for the Hamiltonian. For radiative fields the time-evolution of the wavefunction is described by the time-evolution operator $U(t,t')$ as $\Psi(t)=U(t,t')\Psi(t')$.
Providing $\Delta t=t-t'$ is sufficiently small compared to the characteristic oscillation period of the field perturbation, $U(t,t')$ can be evaluated, for example, using the split-operator method as
\begin{eqnarray}\label{eq:split}
U(t,t') =
\exp\left(\frac{-i\Delta t}{2\hbar}H_0\right)\cdot\exp\left( \frac{-i\Delta t}{\hbar}\sum_{A} E_A((t+t')/2) H_{A}’ \right) \cdot \exp\left(\frac{-i\Delta t}{2\hbar}H_0\right),
\end{eqnarray}
where the elements of the matrices $H_0$ and $H_{A}’$ are given by $\epsilon_{l,J}\delta_{l'l}\delta_{J'J}\delta_{m'm}$ and $\langle l',J',m'|\bar{\mu}_A|l,J,m\rangle$, respectively. The bottleneck operation in Eq.~(\ref{eq:split}) is the evaluation of the matrix exponential at each time step $t=t'+\Delta t/2$, which in our case is computed using the iterative approximation based on Krylov subspace methods, as implemented in the Expokit computational package~\cite{EXPOKIT}. Alternative and more sophisticated representations for the split-operator technique can be found in Refs.~\citenum{Feit82,Bandrauk93}.

\subsection*{Dipole matrix elements}
\label{sec:matelem}

 We aim to derive the relationship between the Einstein A coefficients, $A_{fi}$, and the matrix elements of the electric dipole moment operator, $\langle \psi_f|\bar{\mu}_A|\psi_i\rangle$ ($A=X,Y,Z$), defined relative to the laboratory fixed system. It is convenient to first express $A_{fi}$ in terms of the dipole line strength $S_{fi}$ as
\begin{eqnarray}\label{eq:acoef}
A_{fi} = \frac{64\pi^4}{(4\pi\varepsilon_0)3h}\frac{\tilde{\nu}_{fi}^3}{g_{\rm ns}(2J'+1)}S_{fi},
\end{eqnarray}
where $\tilde{\nu}_{fi}$ is the $f\leftarrow i$ transition frequency in units of cm$^{-1}$, $g_{\rm ns}$ is the nuclear spin statistical weight factor, $J'$ refers to the rotational quantum number of  the final state $f$, and the units of $A_{fi}$ and $S_{fi}$ are $s^{-1}$ and $D^2$, respectively.

 Writing the wavefunctions for the initial and final states as products of rotational and vibrational basis functions, i.e.  $|\psi_i\rangle=|J,k,m\rangle|v\rangle$ and $|\psi_f\rangle=|J',k',m'\rangle|v'\rangle$, the expression for $S_{fi}$ reads~\cite{BunkerBook}
\begin{eqnarray}\label{eq:linestr}
S_{fi} = g_{\rm ns} (2J+1)(2J'+1) \left|(-1)^{k'}\sum_{\sigma=-1}^{1}
\left(\begin{array}{ccc}
J & 1 & J' \\ k & \sigma & -k'
\end{array}\right)
\sum_{\alpha=x,y,z} T_{\sigma\alpha}\langle v' |\bar{\mu}_\alpha| v\rangle
\right|^2,
\end{eqnarray}
where $v$ and $J,k,m$ refer to the vibrational and rotational quanta, respectively, $\mu_\alpha$ ($\alpha=x,y,z$) is the dipole moment in the molecule-fixed axes system, and $T_{\sigma\alpha}$ is the Cartesian-to-spherical tensor transformation matrix
\begin{eqnarray}
\bf{T}=\left[\begin{array}{ccc}
\frac{1}{\sqrt{2}} & -\frac{i}{\sqrt{2}} & 0 \\
0 & 0 & 1 \\
-\frac{1}{\sqrt{2}} & -\frac{i}{\sqrt{2}} & 0 \\
\end{array}\right].
\end{eqnarray}
It should be noted that the actual wavefuncitons $|\psi\rangle$ are expansions in terms of $|J,k,m\rangle|v\rangle$ as basis functions. We omit the expansion coefficients in Eq.~(\ref{eq:linestr}) for the sake of simplicity, which does not alter the methodology presented below.
Using the same variables, the matrix element of the dipole moment operator $\langle \psi_f|\bar{\mu}_A|\psi_i\rangle$ may be expressed as
\begin{eqnarray}\label{eq:mu}
\langle \psi_f|\bar{\mu}_A|\psi_i\rangle &=& \sqrt{(2J+1)(2J'+1)}\left[(-1)^{m'}\sum_{\sigma=-1}^{1} [T^{-1}]_{A\sigma}
\left(\begin{array}{ccc}
J & 1 & J' \\ m & \sigma & -m'
\end{array}\right)\right] \\ \nonumber
&\times&\left[(-1)^{k'}\sum_{\sigma=-1}^{1}
\left(\begin{array}{ccc}
J & 1 & J' \\ k & \sigma & -k'
\end{array}\right)
\sum_{\alpha=x,y,z} T_{\sigma\alpha}\langle v' |\bar{\mu}_\alpha| v\rangle\right].
\end{eqnarray}
By combining Eq.~\eqref{eq:mu} with Eq.~\eqref{eq:acoef} and Eq.~\eqref{eq:linestr}, it is straightforward to write down $\langle \psi_f|\bar{\mu}_A|\psi_i\rangle$ in terms of the quantities $S_{fi}$ and $A_{fi}$ provided by the ExoMol line list. Because of the complex-valued expression under the modulus in Eq.~(\ref{eq:linestr}), an additional number retaining the complex phase must be supplemented for each ro-vibrational transition $f\leftarrow i$ in a line list alongside the value of $A_{fi}$.

 For the sake of simplicity the above formulas were derived using the symmetric top basis functions $|J,k,m\rangle$, which yield complex phases in Eq.~(\ref{eq:linestr}) ranging from 0 to $2\pi$.
In the variational calculations used to generate the ro-vibrational line list, which for polyatomics is performed with the TROVE~\cite{TROVE2007,trove_curv} and RichMol program packages, we employ the basis of symmetrized combinations of the rigid rotor functions $|J,k,\tau, m\rangle = \frac{1}{\sqrt{2}}(|J,k,m\rangle+(-1)^\tau|J,-k,m\rangle)$ for $k>0$ and $|J,0,0, m\rangle$ for $k=0$. Using the symmetrized basis to derive Eqs.~\eqref{eq:linestr} and \eqref{eq:mu}, the expression under the modulus will become purely imaginary, hence the corresponding complex phase factor is $\pm 1$. By defining the complex phase factor as $(-1)^{P_{fi}}$ with $P_{fi}=0$ or 1, the final expression for the dipole matrix elements reads as
\begin{eqnarray}
\langle \psi_f|\bar{\mu}_A|\psi_i\rangle = i\left(\frac{64\pi^4}{(4\pi\varepsilon_0)3h} \frac{\tilde{\nu}_{fi}^3}{(2J'+1)}\right)^{-1/2}
\sqrt{A_{fi}} (-1)^{m'+P_{fi}}\sum_{\sigma=-1}^{1} [T^{-1}]_{A\sigma}
\left(\begin{array}{ccc}
J & 1 & J' \\ m & \sigma & -m'
\end{array}\right).
\end{eqnarray}


\section*{Acknowledgements}

This work was supported by FP7-MC-IEF Project 629237, ERC Advanced Investigator Project 267219, and by STFC Project ST/J002925.
We also thank the DiRAC@Darwin HPC cluster, which is the UK HPC facility for particle physics, astrophysics and cosmology and is supported by STFC and BIS.

\section*{Author contributions statement}

A.Y. conceived and designed the work, and performed the terahertz induced ro-vibrational dynamics simulations. A.O., E.Z. and K.L.C. carried out the analysis related to optoelectrical Sisyphus cooling.  A.O., E.Z., K.L.C., S.N.Y., J.T. and A.Y. all contributed to the writing of the manuscript.

\section*{Additional information}

\textbf{Competing financial interests} The authors declare no competing financial interests. 

\newpage

\begin{table}[!ht]
\tabcolsep=0.2cm
\caption{\label{tab:h2co}Transition frequencies $\nu_{fi}$ (cm$^{-1}$), Einstein A coefficients $A_{fi}$ (s$^{-1}$), and branching ratios $\gamma$ for Sisyphus cooling of H$_2$CO. Ro-vibrational energy levels are labelled as $\ket{\mathrm{Mode},J,K}$ where $0$ refers to the vibrational ground state.}
\begin{center}
	\begin{tabular}{ccccc}
	\hline\hline\\[-3mm]
 	$E\p$ & $E\pp$ & $\nu_{fi}$ & $A_{fi}$ & $\gamma$ \\[0.5mm]
	\hline \\[-3mm]
	$\ket{\nu_1,2,2}$ & $\ket{0,2,2}$ & 2781.79 & 42.4 & 0.66 \\
	$\ket{\nu_1,2,2}$ & $\ket{0,3,2}$ & 2774.50 & 21.1 & 0.33 \\[2mm]
	$\ket{\nu_1,3,3}$ & $\ket{0,3,3}$ & 2781.00 & 47.6 & 0.75 \\
	$\ket{\nu_1,3,3}$ & $\ket{0,4,3}$ & 2771.28 & 15.8 & 0.25 \\[2mm]
	$\ket{\nu_1,4,4}$ & $\ket{0,4,4}$ & 2779.91 & 50.8 & 0.80 \\
	$\ket{\nu_1,4,4}$ & $\ket{0,5,4}$ & 2767.77 & 12.5 & 0.20 \\[2mm]

    $\ket{\nu_2,2,2}$ & $\ket{0,2,2}$ & 1746.01 & 20.9 & 0.67 \\
	$\ket{\nu_2,2,2}$ & $\ket{0,3,2}$ & 1738.73 & 10.4 & 0.33 \\[2mm]	
	$\ket{\nu_2,3,3}$ & $\ket{0,3,3}$ & 1745.99 & 23.6 & 0.75 \\
	$\ket{\nu_2,3,3}$ & $\ket{0,4,3}$ & 1736.27 & 7.8 & 0.25 \\[2mm]
	$\ket{\nu_2,4,4}$ & $\ket{0,4,4}$ & 1745.95 & 25.1 & 0.80 \\
	$\ket{\nu_2,4,4}$ & $\ket{0,5,4}$ & 1733.80 & 6.2 & 0.20 \\
    \hline\hline
    \end{tabular}
\end{center}
\end{table}

\begin{table}[!ht]
\tabcolsep=0.2cm
\caption{\label{tab:ch3cl}Transition frequencies $\nu_{fi}$ (cm$^{-1}$), Einstein A coefficients $A_{fi}$ (s$^{-1}$) and branching ratios $\gamma$ for Sisyphus cooling of CH$_3$Cl. Ro-vibrational energy levels are labelled as $\ket{\mathrm{Mode},J,K}$ where $0$ refers to the vibrational ground state.}
\begin{center}
	\begin{tabular}{ccccc}
	\hline\hline\\[-3mm]
 	$E\p$ & $E\pp$ & $\nu_{fi}$ & $A_{fi}$ & $\gamma$ \\[0.5mm]
	\hline\\[-3mm]
	$\ket{\nu_1,2,2}$ & $\ket{0,2,2}$ & 2968.89 & 15.3 & 0.65 \\
	$\ket{\nu_1,2,2}$ & $\ket{0,3,2}$ & 2966.23 & 7.6 & 0.32 \\[2mm]
	$\ket{\nu_1,3,3}$ & $\ket{0,3,3}$ & 2968.61 & 17.3 & 0.73 \\
	$\ket{\nu_1,3,3}$ & $\ket{0,4,3}$ & 2965.07 & 5.7 & 0.24 \\[2mm]
	$\ket{\nu_1,4,4}$ & $\ket{0,4,4}$ & 2968.23 & 18.4 & 0.78 \\
	$\ket{\nu_1,4,4}$ & $\ket{0,5,4}$ & 2963.80 & 4.6 & 0.19 \\
    \hline\hline
    \end{tabular}
\end{center}
\end{table}

\newpage

\newpage

\end{document}